\documentclass[prl,letterpaper,onecolumn,preprint,showpacs]{revtex4}

\newcommand{\sisio}{Si-SiO$_{2}$ }

\newcommand{\Phos}{$^{31}$P }
\newcommand{\Dm}{D$^{-}$ }

\newcommand{\DI}{$\Delta I$ }
\newcommand{\Toe}{T$_{1}^{e}$ }

\newcommand{\Ton}{T$_{1}^{n}$ }
\newcommand{\Ttn}{T$_{2}^{n}$ }

\usepackage{times}
\usepackage{graphicx}% Include figure files
\usepackage{dcolumn}% Align table columns on decimal point
\usepackage{bm}% bold math
\bibpunct{}{}{,}{s}{}{}
\begin{document}

\title{Electrically detected spin echoes of donor nuclei in silicon}

\author{D. R. McCamey}\email{dane.mccamey@sydney.edu.au}
\affiliation{School of Physics, University of Sydney, NSW 2006, Australia}

\author{C. Boehme}
\affiliation{Department of Physics, University of Utah, 115 South
1400 East Rm 201, Salt Lake City, Utah 84112}

\author{G. W. Morley}
\affiliation{London Centre for Nanotechnology and Department of Physics and Astronomy, University College London, 17-19 Gordon Street, London WC1H 0AH, United Kingdom}

\author{J. van Tol}
\affiliation{Center for Interdisciplinary Magnetic Resonance,
National High Magnetic Field Laboratory at Florida State University,
Tallahassee, Florida 32310, USA}

\date{\today}

%--------------------------------------------------------------------
%ABSTRACT
%--------------------------------------------------------------------

\begin{abstract}

The ability to probe the spin properties of solid state systems electrically underlies a wide variety of emerging technology. Here, we extend electrical readout of the nuclear spin states of phosphorus donors in silicon to the coherent regime with modified Hahn echo sequences. We find that, whilst the nuclear spins have electrically detected phase coherence times exceeding 2 ms, they are nonetheless limited by the artificially shortened lifetime of the probing donor electron.

\end{abstract}


%
%76. Magnetic resonances and relaxations in condensed matter,
%M�ssbauer effect 76.30.-v Electron paramagnetic resonance and
%relaxation (see also 76.30.Da Ions and impurities: general

%72.10.Fk	Scattering by point defects, dislocations, surfaces, and other imperfections (including Kondo effect)

%\keywords{}%Use showkeys class option if keyword
\maketitle

%--------------------------------------------------------------------
%INTRODUCTION
%--------------------------------------------------------------------

Using spin to encode and process information lies at the heart of a range of emerging technologies\cite{Wolf2001,Zutic2004,Kane1998,Dediu2009,Ladd2010}. Whilst electron spins are an obvious choice for manipulating information, nuclear spins provide a robust system in which to store spin information for long periods of time\cite{Morton2008}. However, whilst a number of optical\cite{Neumann2010,Steger2011} and quantum Hall based techniques \cite{Yusa2005} exist for reading the state of small ensembles of nuclear spins, until recently there has been no technique for doing so which is compatible with conventional electronic devices, particularly those in silicon. Using pulsed spin resonance, we recently demonstrated the ability to electrically measure the nuclear spin state of phosphorus donors in silicon\cite{McCamey2010a} with a long spin lifetime. However, whilst the nuclear spin lifetime (\Ton) imposes a limit on the storage of classical information, it is the nuclear spin phase coherence time (\Ttn) which sets a limit on the storage of quantum information, usually shorter than the spin lifetime. In this Letter, we utilize a spin echo technique to show that electrical readout of donor nuclear spins is compatible with phase coherence times \Ttn exceeding 2 ms, nearly two orders of magnitude longer than that previously seen\cite{Hoehne2011a}. We also determine that the artificially shortened donor electron lifetime \Toe limits \Ttn via hyperfine coupling to the nuclei \cite{Morton2008}, and discuss ways to overcome this limitation.

To access the nuclear spin state of phosphorus donors in silicon, we exploit the precise hyperfine coupling between the donor nucleus and electron. By selectively exciting the donor electron dependent on the state of the nuclear spin, we obtain a sensitive probe of the nuclear spin \cite{Sarovar2008, McCamey2010a}. A variety of methods can then be used to readout the selective excitation of the electron. For example, coupling the donor electron to the island of a nearby single electron transistor can allow single spin readout, but this technique requires the donors to be near an interface \cite{Morello2010}. Whilst this technique has not yet been used to readout coherent states of the electron, we anticipate the possibility that proximity of defects at the \sisio interface will result in a reduced electron phase coherence (in the 1-10 $\mu$s range) \cite{Schenkel2005, Sousa2007, Paik2010}. Utilizing spin dependent recombination also results in reduced electron coherence times\cite{Huebl2008,Paik2010}. However, Hoehne \emph{et al.}\cite{Hoehne2011a} have recently shown that electrical readout of coherent nuclear spin motion could be detected  with this approach, with a phase coherence time \Ttn $\sim 50 \mu$s. 

\begin{figure}[h!]
\centering\includegraphics[width=6.5cm]{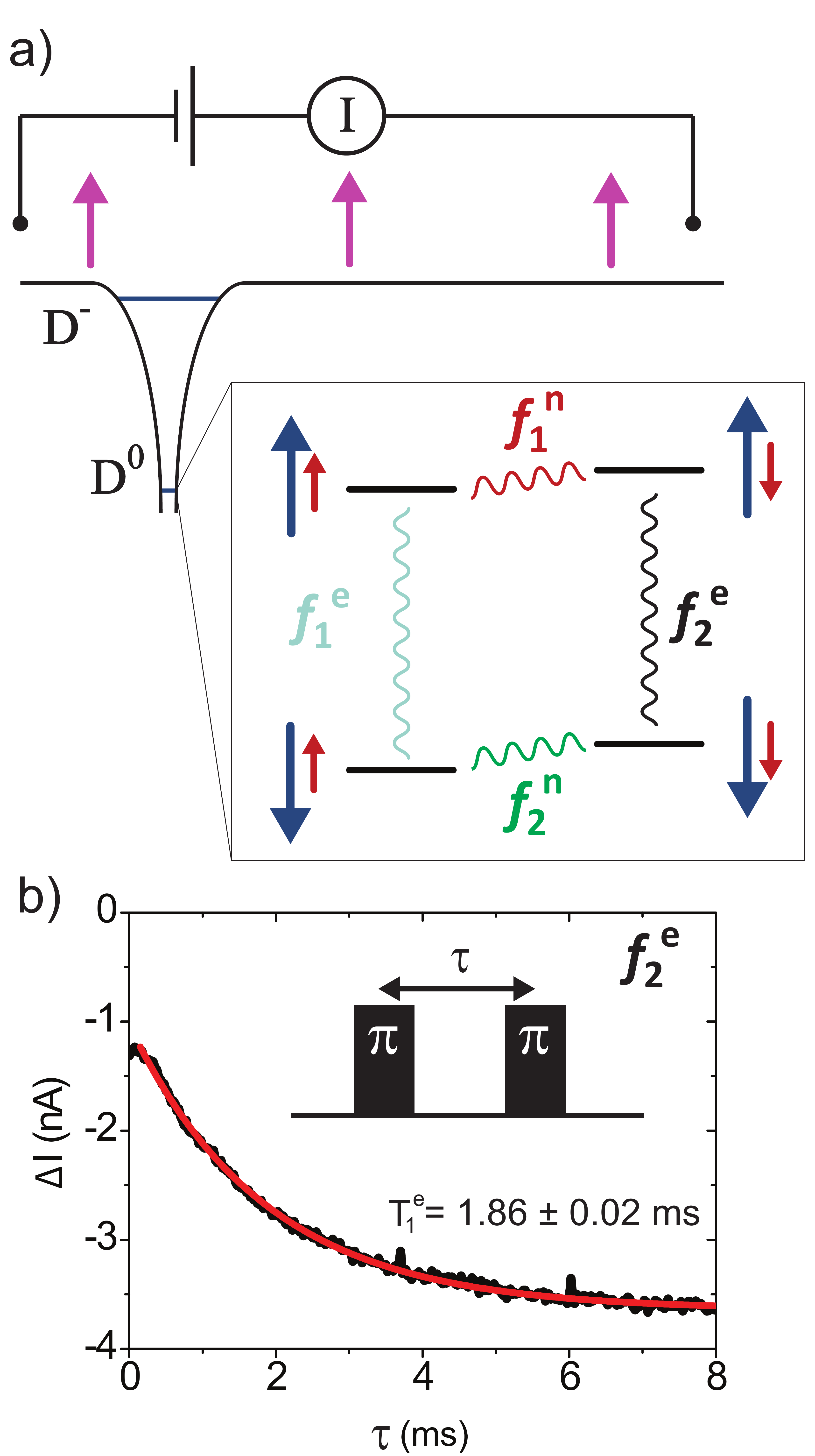}
\caption{\label{fig:SiP} a) The spin state of the P donor electron in silicon is detected by monitoring the photocurrent. Electrons from the conduction band are trapped in the \Dm state only when they can form a singlet with the donor electron. The four eigenstates of the Si:P donor system are shown with large blue arrows for the (spin 1/2) electron spin and smaller red arrows for the (spin 1/2) nuclear spin. The hyperfine interaction betwen the nucleus and electron lifts the degeneracy for the four transitions which can be driven using spin resonance. The nuclear spin state thus determines if resonance (and thus a change in the photocurrent) occurs at $f_{1}^{e}$ or $f_{2}^{e}$. Probing one of these transitions therefore allows the nuclear spin state to be determined. For the experiments reported here, $f_{2}^{e} = 240$ GHz, $f_{1}^{n} = 88.60$ MHz, and $f_{2}^{n} = 201.24$ MHz. b) The change in photocurrent following an inversion-recovery type experiment (inset) on the nuclear spin down transition ($f_{2}^{e}$). The lifetime of the electron spin \Toe$= 1.86 \pm 0.02$ ms.}
\end{figure}

Here, we utilize spin-dependent trapping of photoexcited electrons into the \Dm state of the \Phos donor \cite{Morley2008a,Thornton1973} [Fig. \ref{fig:SiP}(a)]. We have previously used this mechanism to readout electron spin states with coherence times exceeding 150 $\mu$s \cite{Morley2008a}, and to determine the state of the nuclear spin following transfer and classical storage of the electron spin state\cite{McCamey2010a}. The donors we measure are not restricted to the interface, and are therefore able to exhibit much longer coherence times. The electron coherence time is in this case limited by the trapping of conduction electrons into the \Dm state, a process which should be controllable by modifying the carrier density. The electron spin lifetime is determined by the occupation time of the electron \cite{Morley2008a}. Figure \ref{fig:SiP}(b) shows the result of an electrically detected inversion recovery experiment which allows us to determine the spin lifetime of the donor electrons; in these measurements \Toe $= 1.86 \pm 0.02$ ms.

In our earlier work, we were unable to demonstrate coherent nuclear spin manipulation due to the spatial inhomogeneity of the rf radiation used to drive nuclear spin transitions. However, by utilising spin echo techniques, we now show that electrically detected nuclear spin phase coherence times can exceed 2 ms, nearly two orders of magnitude longer than previously demonstrated in silicon. 

\emph{Experiment-}
A phosphorus doped silicon ($n$[$^{31}$P] $= 10^{15}$ cm$^{-3}$) sample, described in detail in reference \onlinecite{McCamey2010a}, was used for these experiments. Electrical contacts to the silicon are made with thin film aluminum contacts. White light is shone onto the sample, and a constant current source with a slow time constant is used to provide a quasistatic photocurrent, $I = 200$ nA. Transient changes in this current, $\Delta I(t)$ are amplified and recorded with a digital oscilloscope. All experiments are performed at a temperature of 3.8 K. Electron spin transitions are driven using a custom built spectrometer which operates at 240 GHz \cite{Tol2005,Morley2008}, and which also contains an NMR coil to drive the nuclear spin transitions ($f_{1}^{n} =$ 88.60 MHz, $f_{2}^{n} =$ 201.24 MHz  at $B =$ 8.56370 T).

\begin{figure}
\centering\includegraphics[width=8.5cm]{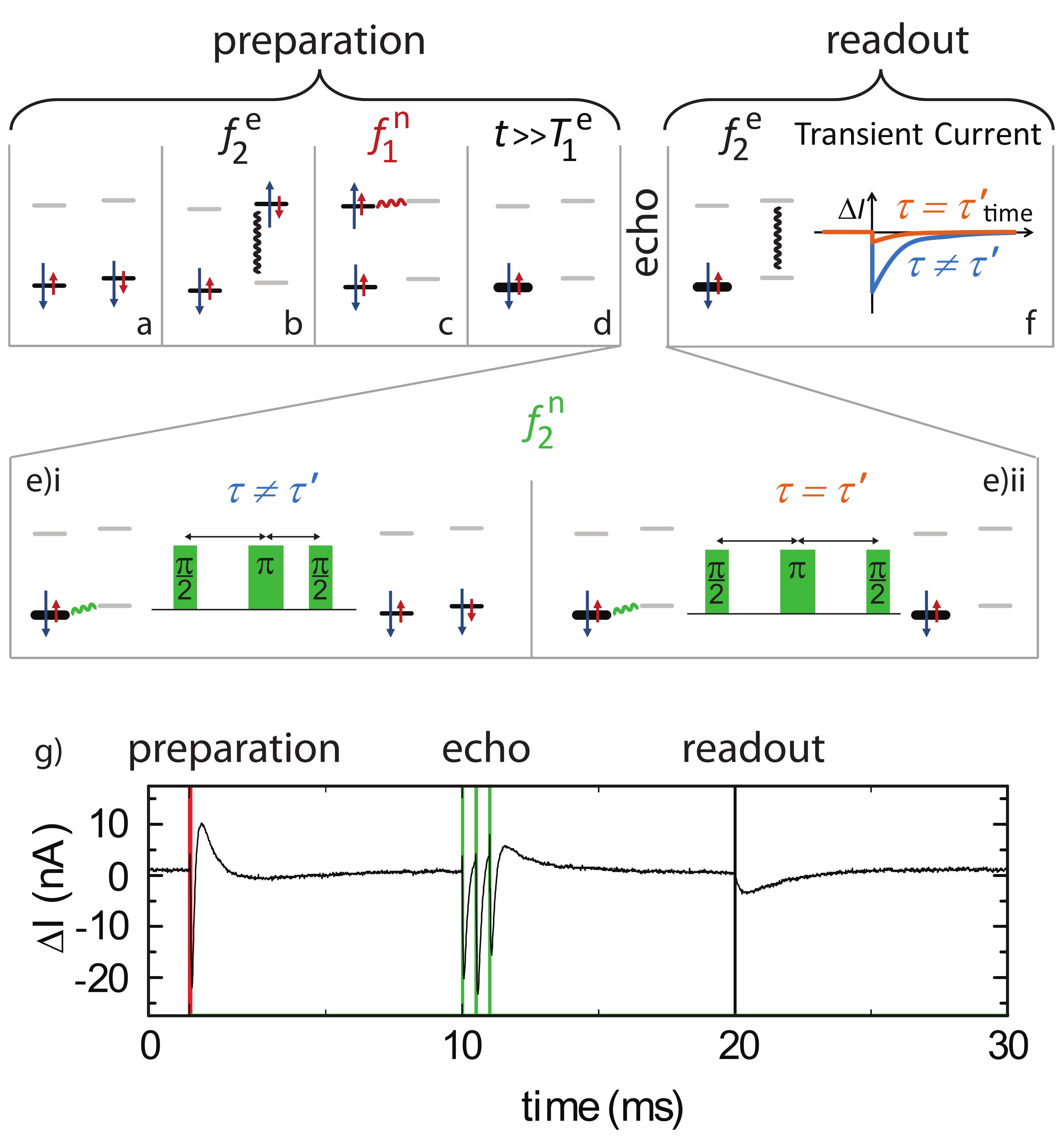}
\caption{\label{fig:echo} Pulse sequence for observing nuclear spin echoes. (a-d) Preparation of a polarized nuclear spin population is obtained by mapping the large electron spin polarization to the nuclear spins with a two pulse sequence. (e) The nuclear spin echo is performed on the electron spin down manifold. (f) Readout of the resulting nuclear spin population is achieved by selectively exciting the electron spins corresponding to nuclear spin down, and monitoring the current change throught the sample after the excitation. g) The change in current through the sample during an entire pulse sequence, with $\tau = 0.5$ ms.}
\end{figure}

To obtain a starting point for our measurments, we first polarize the nuclear spin [Fig. \ref{fig:echo}(a)-(d)]. Whilst this could be achieved in a number of ways\cite{McCamey2009,Tol2009}, we make use of the fact that the donor electrons are nearly entirely polarized by implementing a simple swap pulse sequence  ($\pi (f_{2}^{e})$ - $\pi (f_{1}^{n})$), similar to the sequence used to store classical information in earlier work\cite{McCamey2010a,Simmons2011},which results in an excess of nuclear spin up. 

Following nuclear spin polarization, the system is allowed to recover for $t_{\mathrm{wait}} = 10$ ms, long enough for the electrons to relax to the ground state ($t_{\mathrm{wait}} >$ \Toe), but still much shorter than the nuclear spin lifetime ($t_{\mathrm{wait}} <$ \Ton) so that the polarization is maintained. A Hahn echo is then performed on the $f_{2}^{n}$ transition, consisting of a $\pi/2 - \tau - \pi - \tau^{\prime} - \pi/2$ pulse sequence. A perfect echo should leave the nuclear spin polarization unchanged [Fig. \ref{fig:echo}(e)ii], whereas a non-optimal echo sequence (e.g. with $\tau \neq \tau^{\prime}$) should result in a decrease in the polarization, as some nuclear spin population returns to the spin down state [Fig. \ref{fig:echo}(e)i]. To determine the resulting nuclear spin polarization, a readout pulse, $\pi (f_{2}^{e})$, is applied $10$ ms after the start of the echo sequence. The magnitude of the change in current following the readout pulse is proportional to the nuclear spin down population. 

Figure \ref{fig:echo}(g) shows the current through the sample during one such polarization-echo-readout sequence. The transient behaviour at 0 and 10 ms results from the application of the RF radiation used to drive nuclear spin transitions. It is non-resonant, and most likely due to resistance changes from heating the sample. The current change at 20 ms is due to the electron trapping described above, and only occurs when the electron spin resonance conditions are matched (in this case for the  $f_{2}^{e}$ transition).

\begin{figure}
\centering\includegraphics[width=7cm]{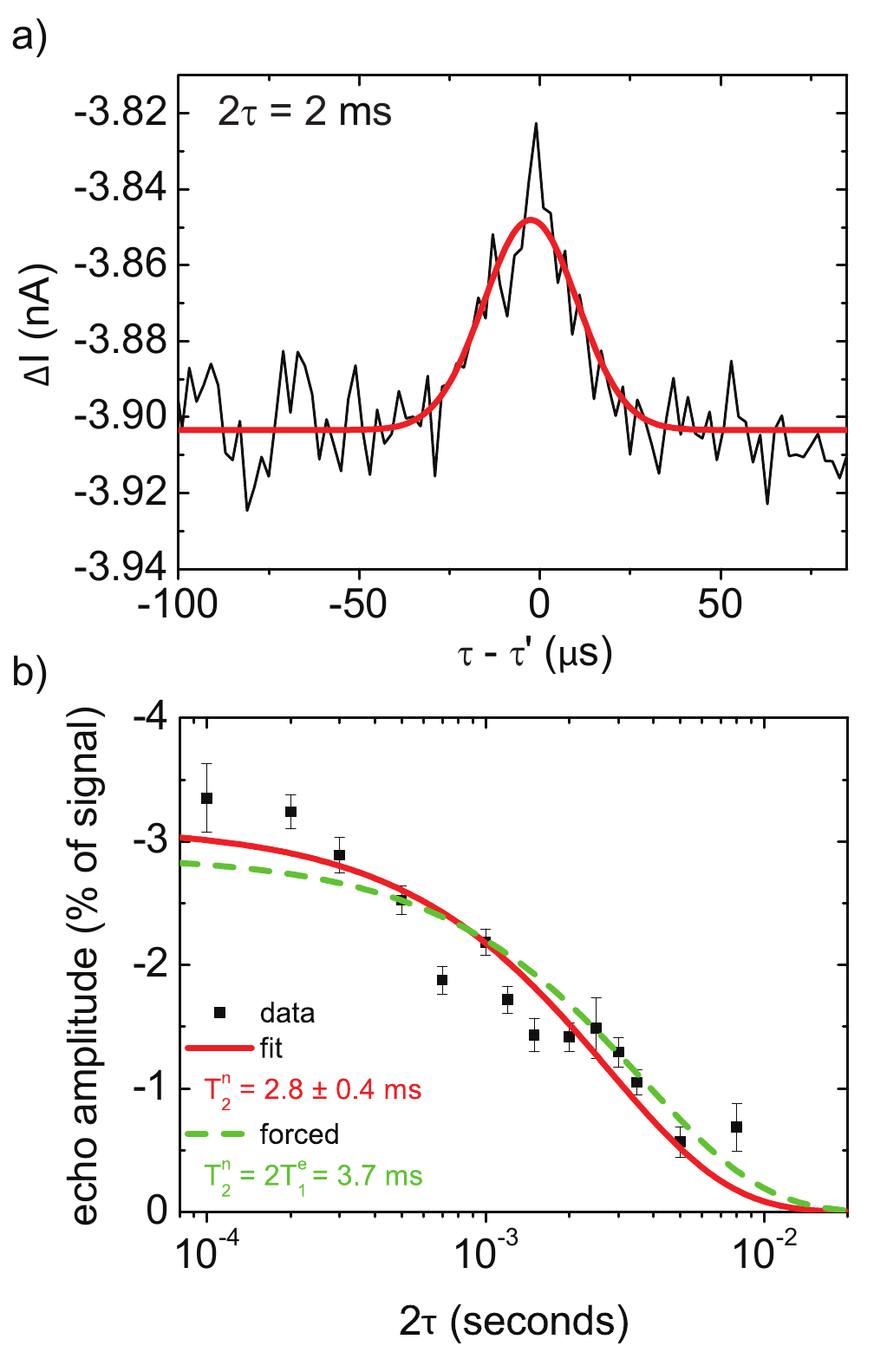}
\caption{\label{fig:echoamp} Nuclear echoes. a) The change in photocurrent following the readout pulse for an echo sequence with $\tau = 1$ ms. A clear echo is seen when $\tau = \tau^{\prime}$. b) The amplitude of the echo is plotted for a range of $\tau$. The data are fit with a simple exponentially decaying function, yielding a nuclear spin coherence time \Ttn $= 2.8 \pm 0.4$ ms (red solid line). The data are fit nearly as well if \Ttn is fixed to be 2\Toe (green dashed line). Note the logarithmic scale on the x-axis.}
\end{figure}

To observe the spin echo, the transient current, $\Delta I$, following the readout pulse is measured\footnote{We average the change in current for 300 $\mu$s following the readout pulse} as a function of $\tau - \tau^{\prime}$, for a fixed $\tau$. Figure \ref{fig:echoamp}(a) shows \DI during an echo sequence with $\tau = 1$ ms.  We expect rephasing to occur when $\tau - \tau^{\prime} = 0$, and an echo is indeed observed when this condition is satisifed. The direction of the signal change (i.e. \DI becomes smaller) is as expected. We note that the magnitude of the signal does not go to zero at the peak of the echo, and we attribute this to the non-ideal pulses used to drive both electronic and nuclear spin transitions. 

To determine the phase coherence of the nuclear spin system, the echo sequence was repeated for a range of $\tau$. Figure \ref{fig:echoamp}(b) shows the echo amplitude $\Delta I (\tau = \tau^{\prime})-\Delta I (\tau \neq \tau^{\prime})$ as a percentage of $\Delta~I (\tau \neq \tau^{\prime}$), obtained by fitting each echo with a Gaussian peak function. The data are well fit by a simple exponential decay with a time constant \Ttn $= 2.8 \pm 0.4$ ms. This represents an increase of nearly 2 orders of magnitude for electrical readout of nuclear spin phase coherence of donors in silicon, and is to our knowledge the longest electrically measured spin phase coherence time in any condensed matter system\cite{Yusa2005}.

We note that, whilst long, the \Ttn we measure is nonetheless shorter than has been previously measured for donors in silicon using conventional means\cite{Morton2008}. A number of processes exist which limit \Ttn in those measurements, but the fundamental limit seems to be the lifetime of the hyperfine coupled donor electron\cite{Morton2008,Tol2009}: \Ttn $\leq 2$\Toe.  In the experiments reported here, the nuclear coherence time measured is indeed very close to twice the spin lifetime of the coupled donor electron, \Ttn $=2.8 \pm 0.4$ ms $ \approx 2$\Toe $= 3.72 \pm 0.04$ ms [dashed line, Fig. \ref{fig:echoamp}(b)], indicating that this process is also the dominant dephasing mechanism in this work. However, it is important to note that the electron spins in these experiments have artificially shortened lifetimes due to their interaction with the photoexcited conduction electrons required for readout, and as such we anticipate that increasing \Toe should lead to longer \Ttn. A number of modifications to the experiment described here would help in this regard. A pulsed photoexcitation scheme could be utilized to enable readout only when required, retaining longer coherence times useful for computing and storage otherwise. Utilizing a MOSFET structure\cite{lo07,Beveren2008,Lo2011} would also allow this modification, as well as providing more accurate control of the donor density (and thus scattering time between free and donor electrons) which could be used to tune the nuclear spin lifetime \cite{Stemeroff2011}.

To summarize, we have demonstrated electrically measured spin echoes from phosphorus donor nuclear spins in silicon. We find a spin phase coherence time \Ttn $=2.8 \pm 0.4$ ms, the longest electrically measured spin phase coherence to date. Evidence suggests that \Ttn is limited by the lifetime of the hyperfine coupled donor electron, which bodes well for increasing this time by utilizing methods to reduce the free electron density when reaodut is not required, for example by utilizing pulsed photoexcitation or MOSFET style devices. Finally, we note that, as well as the relevance to technological devices based on spin, the techniques used here may find applications as a tool for investigating the physics of systems comprised of a small number of spins, where conventional spin resonance techniques are not suitable.

%--------------------------------------------------------------------
%ACKNOWLEDGEMENTS
%--------------------------------------------------------------------

This work was supported by the Australian Research Council (DP1093526). DRM acknowledges an ARC Postdoctoral Fellowship. CB acknowledges support of the NSF through a Career Award (0953225). GWM acknowledges an 1851 Research Fellowship and support from the EPSRC COMPASSS grant. A portion of this work was performed at, and supported by a Visiting Scientist Program Grant from, the National High Magnetic Field Laboratory, which is supported by National Science Foundation Cooperative Agreement No. DMR-0654118, the State of Florida, and the U.S. Department of Energy.

%\nocite{*}
%\bibliography{nuclear_echo}

\newpage
%--------------------------------------------------------------------
%FIGURE
%--------------------------------------------------------------------

%--------------------------------------------------------------------
%FIGURE2
%--------------------------------------------------------------------

%--------------------------------------------------------------------
%FIGURE3
%--------------------------------------------------------------------

\end{document}